\documentclass[aps,prb,twocolumn,showpacs,superscriptaddress,floatfix]{revtex4-1}
\pdfoutput=1
\usepackage{amsmath,amssymb}
\usepackage{graphicx}
\usepackage{color}
\usepackage{rotating}
\usepackage{bm}
\usepackage{setspace}
\usepackage{hyperref}
\usepackage{float}
\usepackage{soul}
\usepackage{color}

\usepackage{xcolor}

\usepackage{enumerate}

\begin{document}

\title{Relaxation and Domain Formation in Incommensurate 2D Heterostructures}
\author{Stephen Carr}
	\affiliation{Department of Physics, Harvard University, Cambridge, Massachusetts 02138, USA.}
\author{Daniel Massatt}
	\affiliation{School of Mathematics, University of Minnesota, Minneapolis, Minnesota, 55455, USA.}
\author{Steven B. Torrisi}
	\affiliation{Department of Physics, Harvard University, Cambridge, Massachusetts 02138, USA.}
\author{Paul Cazeaux}
	\affiliation{Department of Mathematics, University of Kansas, Lawrence, Kansas, 66045, USA.}
\author{Mitchell Luskin}
	\affiliation{School of Mathematics, University of Minnesota, Minneapolis, Minnesota, 55455, USA.}
\author{Efthimios Kaxiras}
	\affiliation{Department of Physics, Harvard University, Cambridge, Massachusetts 02138, USA.}
	\affiliation{John A. Paulson School of Engineering and Applied Sciences, Harvard University, Cambridge, Massachusetts 02138, USA.}
\date{\today}                                        
\begin{abstract}

We introduce configuration space as a natural representation for calculating the mechanical relaxation patterns of incommensurate two-dimensional (2D) bilayers, bypassing supercell approximations to encompass aperiodic relaxation patterns.
The approach can be applied to a wide variety of 2D materials through the use of a continuum model in combination with a generalized stacking fault energy for interlayer interactions.
We present computational results for small-angle twisted bilayer graphene and molybdenum disulfide (MoS$_2$), a representative material of the transition metal dichalcogenide (TMDC) family of 2D semiconductors.
We calculate accurate relaxations for MoS$_2$ even at small twist-angle values, enabled by the fact that our approach does not rely on empirical atomistic potentials for interlayer coupling.
The results demonstrate the efficiency of the configuration space method by computing relaxations with minimal computational cost for twist angles down to $0.05^\circ$, which is smaller than what can be explored by any available real space techniques.
We also outline a general explanation of domain formation in 2D bilayers with nearly-aligned lattices, taking advantage of the relationship between real space and configuration space.

\end{abstract}

\maketitle

\newcommand{\E}{\mathcal{E}}
\newcommand{\R}{\mathcal{R}}
\newcommand{\Z}{\mathbb{Z}^2}
\newcommand{\inter}{\text{inter}}
\newcommand{\intra}{\text{intra}}
\newcommand{\GSFE}{V_{\text{GSFE}}}
\def\Xint#1{\mathchoice
    {\XXint\displaystyle\textstyle{#1}}%
    {\XXint\textstyle\scriptstyle{#1}}%
    {\XXint\scriptstyle\scriptscriptstyle{#1}}%
    {\XXint\scriptscriptstyle\scriptscriptstyle{#1}}%
      \!\int}
\def\XXint#1#2#3{{\setbox0=\hbox{$#1{#2#3}{\int}$}
    \vcenter{\hbox{$#2#3$}}\kern-.5\wd0}}
\def\dashint{\Xint-}

Layered materials consist of 2D atomically thin sheets which are weakly coupled by the van der Waals force.
For understanding the electronic and mechanical properties of multilayered structures of such materials, it is useful to view them as a series of conventional crystals with a weak perturbative interaction between sheets \cite{Tritsaris2016}.
Bilayer systems with slight lattice misalignment due to differing lattice constants or relative twist-angle are of interest in optical and transport experiments \cite{Dean2013, Hunt2013, Geim2013,  Zhang2017}.
In small-angle twisted bilayer graphene (tBLG), highly regular domain-wall patterns have been observed experimentally and studied theoretically \cite{Alden2013, Nam2017, Zhang2018}
The appearance of domain walls is the result of atomic relaxation which serves to minimize the additional energy due to misalignment.
Under electric-field gating the domain walls give rise to interesting topologically-protected edge states \cite{Vaezi2013, Yin2016, Jiang2017}.
Understanding this relaxation and predicting its behavior in other nearly-aligned bilayers may be useful in the search for topological edge states and quantum information applications.

To this end, we chose to study three different bilayer systems, graphene and the two high-symmetry alignments of MoS$_2$, which is a standard representative of the transition-metal dichalcogenide family of 2D materials.
A unit-cell with basis vectors $\mathbf{a_1} = a (1,0)$ and $\mathbf{a_2} = a (\sqrt{3}/2,1/2)$ is used, where the lattice parameter $a$ for graphene is $2.47$ \AA\; and for MoS$_2$ is $3.18$ \AA\;.
Insight into the mechanical domain-wall formation can be gained by paying special attention to the relationship between intralayer bonding energies and interlayer stacking energies, the latter arising from the much weaker van der Waals force.
To do this, a consistent model must be chosen for both types of energy.
We will assume smooth and slowly-varying relaxation in each layer, described by a position-dependent displacement vector field $\mathbf{U}_i(r)$ where $i$ indexes the layer number.
We will only consider in-plane relaxation, which is appropriate for a bilayer encapsulated in a stiff substrate, although the method can be extended for out-of-plane relaxations as well.
Such encapsulated systems show improved optical and electronic transport properties \cite{Dean2010} and are of great experimental interest.

Under these assumptions, the intralayer energy for a layer is well described by a linear isotropic continuum approximation

\begin{align} \label{eq:intralayer_E}
	E_\textrm{intra} &= \lim_{R \to \infty} \frac{1}{|B_R|} \int_{B_R} \frac{1}{2} \mathcal{E} \left( \nabla \mathbf{U}_i \right) \; C_i \; \mathcal{E} \left( \nabla \mathbf{U}_i \right) dr \nonumber\\
	&= \lim_{R \to \infty} \frac{1}{|B_R|} \int_{B_R} \frac{1}{2} \bigg[ G_i (\partial_x U_{ix} + \partial_y U_{iy})^2 + \\
	& \quad K_i ( (\partial_x U_{ix} - \partial_y U_{iy})^2 + (\partial_x U_{iy} + \partial_y U_{ix})^2) \bigg] dr \nonumber
\end{align}

where $\mathcal{E} (\nabla \mathbf{U}_i ) = \frac{1}{2} (\nabla \mathbf{U}_i + \nabla \mathbf{U}_i^T)$ is the $2 \times 2$ infinitesimal strain tensor and $B_R$ is a sphere of radius $R$ used to normalize the integral.
The fourth-order stiffness tensor $C$ depends on the two parameters $G$ (bulk modulus) and $K$ (shear modulus) which represent the energy cost associated with strain.
This approximation does not capture short-range symmetry-breaking effects like Peierls distortions, but can describe the long-range domain-walls observed in twisted bilayer graphene.

For the interlayer energy, we use the generalized stacking fault energy (GSFE) surface. This concept was originally used to describe the energy of defects in bulk crystals \cite{Kaxiras1993,Rice1994,Waghmare1999,Waghmare2000}
and has recently been employed for explaining relaxation in graphene and hBN bilayers \cite{Zhou2015,Dai2016}.
The GSFE provides the interlayer energy per unit cell and depends only on the relative stacking between two successive layers.
We denote this functional by $\GSFE$, with the initial local stacking configuration between layers given by the 2D vector $\mathbf{b}(\mathbf{r})$.
We can obtain the stacking after relaxation by adding in the displacement fields $\mathbf{U}_i(\mathbf{r})$, giving a normalized interlayer energy:

\begin{equation} \label{interlayer_E}
	E_\textrm{inter} = \lim_{R \to \infty} \frac{1}{|B_R|} \int_{B_R} \GSFE (\mathbf{b}(\mathbf{r}) + \mathbf{U}_1(\mathbf{r}) - \mathbf{U}_2(\mathbf{r})) d\mathbf{r}.
\end{equation}

These intralayer and interlayer couplings are obtained from total-energy calculations based on density-functional theory (DFT) with the Vienna Ab initio Simulation Package (VASP) \cite{Kresse1993, Kresse1996}.
For the intralayer coupling, isotropic and anisotropic strain are applied to an optimized monolayer unit cell, distorting the $x$ and $y$ axes by $\pm 1.5 \%$ in steps of $0.3\%$.
We obtain $G$ and $K$ of Eq. \eqref{eq:intralayer_E} by linear fitting of the ground-state energy dependence on this applied strain.
For MoS$_2$, the sulfur atom heights for each strain sample are relaxed while calculating the ground-state energy.

For the interlayer GSFE, we use previously reported DFT results for bilayer graphene stacking \cite{Zhou2015}.
In MoS$_2$, the GSFE was parameterized by evaluating the energy on a grid of points for an MoS$_2$ bilayer, with the van der Waals force implemented through the vdW-DFT method using the SCAN+rVV10 functional \cite{Klimes2011, Peng2016}.
The in-plane positions of all atoms in the bilayer are fixed but they are allowed to relax in the out-of-plane direction.
The top layer is shifted relative to the bottom layer over a $9 \times 9$ grid in the unit cell to sample the GSFE energy landscape.
To fit the $\GSFE$ to this set of values, we use a form similar to that used by Zhou et. al. \cite{Zhou2015}, but with modifications that better highlight how the symmetry of the bilayer affects the GSFE.
First, we define two parameters, $(v, w) \in [0,2\pi] \times [0,2\pi]$, which describe $\mathbf{b}(\mathbf{r})$ in terms of the unit-cell vectors. For the bilayers studied here, $v$ and $w$ are related to the stacking vector $(b_x$, $b_y)$ by

\begin{equation}
	\begin{pmatrix}
		v \\
		w
	\end{pmatrix}
 	= \frac{2\pi}{\alpha}
 	\begin{bmatrix}
 		1 & -1/\sqrt{3} \\
 		0 &  2/\sqrt{3}
	\end{bmatrix}
	\begin{pmatrix}
 		b_x \\
 		b_y
	\end{pmatrix}.
\end{equation}

The GSFE can then be written in a relatively simple form in the $(v,w)$ basis:

\begin{align}
	\GSFE = c_0		&+ c_1 (\cos{v} + \cos{w} + \cos{(v+w)}) \nonumber \\
    				&+ c_2 (\cos(v + 2 w) + \cos(v - w) + \cos(2 v + w)) \nonumber\\
    				&+ c_3 (\cos(2 v) + \cos(2 w) + \cos(2 v + 2 w)) \nonumber\\
   					&+ c_4 (\sin{v} + \sin{w} - \sin(v+w)) \\
    				&+ c_5 (\sin(2 v + 2 w) - \sin(2 v) - \sin(2 w)) \nonumber
\end{align}

with the coefficients $c_0, \dots, c_5$ given in Table \ref{table:gsfe_coeffs}. For hexagonal systems like graphene that have symmetry between the AB and BA stackings, the coefficients of the sine terms are constrained to be zero as the GSFE functional must be even around the origin (AA stacking).

\begin{table*}
\begin{center}
  \begin{tabular}{| l || r | r || r | r | r | r | r | r |}
    \hline
    Material 				& $G$ & $K$ & $c_0$ & $c_1$ & $c_2$ & $c_3$ & $c_4$ & $c_5$ \\ \hline
    Graphene 				& 69,518 & 47,352 & 6.832 & 4.064 & -0.374 & -0.095 & 0.000 & 0.000 \\ \hline
    $0^\circ$ MoS$_2$ 		& 49,866 & 31,548 & 27.332 & 14.02 & -2.542 & -0.884 & 0.000 & 0.000 \\ \hline
    $180^\circ$ MoS$_2$ 	& 49,866 & 31,548 & 30.423 & 12.322 & -2.077 & -0.783 & 2.397 & 0.259 \\
    \hline
  \end{tabular}
  \caption{Coefficients for the strain energy and Fourier components of the GSFE in bilayer graphene and the two high-symmetry forms of bilayer MoS$_2$. All values are in units of meV per unit-cell.}
\label{table:gsfe_coeffs}
\end{center}
\end{table*}

In the supercell case, the large body limits in Eq.s \eqref{eq:intralayer_E} and \eqref{interlayer_E}  become a normalized integral over a supercell. For an incommensurate bilayer, $\mathbf{U}_i(\mathbf{r})$ are aperiodic and so a different formalism is required. It should rely only on the local stacking arrangement between layers, not a periodic real space structure. The collection of all local stackings in an incommensurate system forms a dense compact domain called configuration space \cite{Massatt2017, Cazeaux2017, Cances2017}, and addresses this problem.

\begin{figure}
	\includegraphics[width=\linewidth]{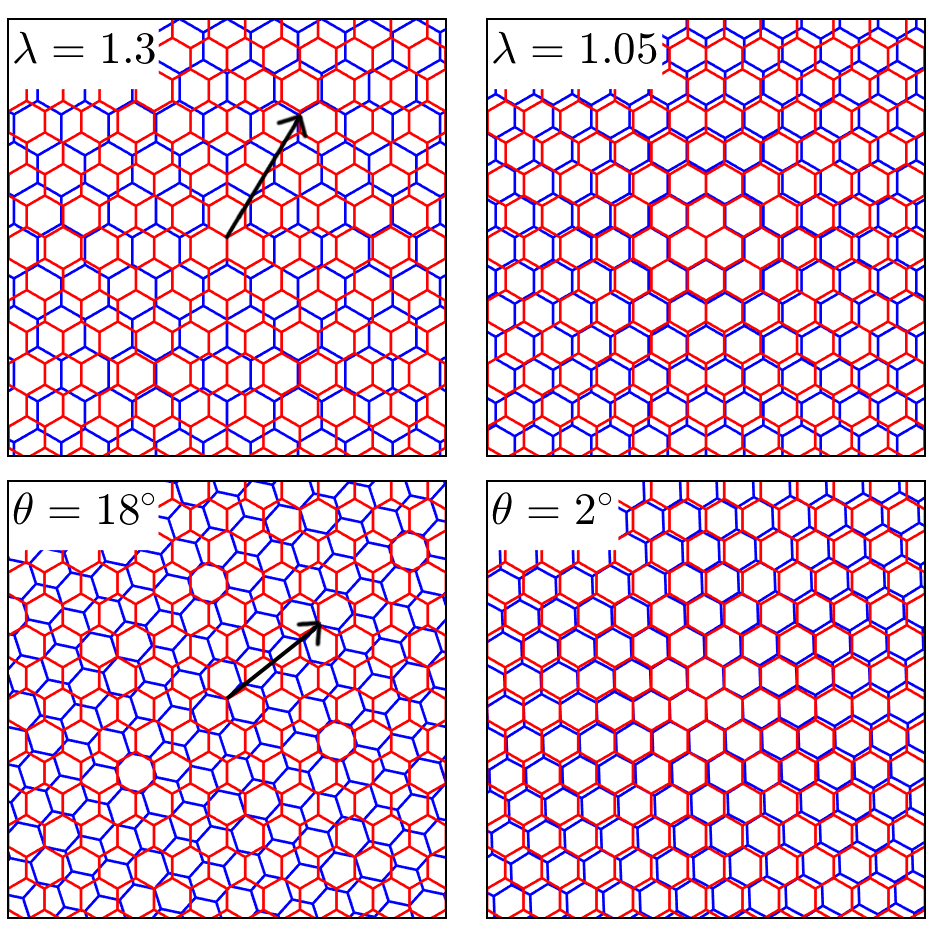}
	\caption{
	Examples of misaligned lattices. 
	Top panels show bilayers whose lattice constants differ by factor of $\lambda$ while the bottom panels show bilayers with relative twist angle $\theta$.
	The left panels show misaligned structures with small moir\'e length (moir\'e wavelength given by the black arrow), while the right panels show nearly-aligned structures with large moir\'e length.}
	\label{fig:lattice_alignments}
\end{figure}

\begin{figure}
	\includegraphics[width=\linewidth]{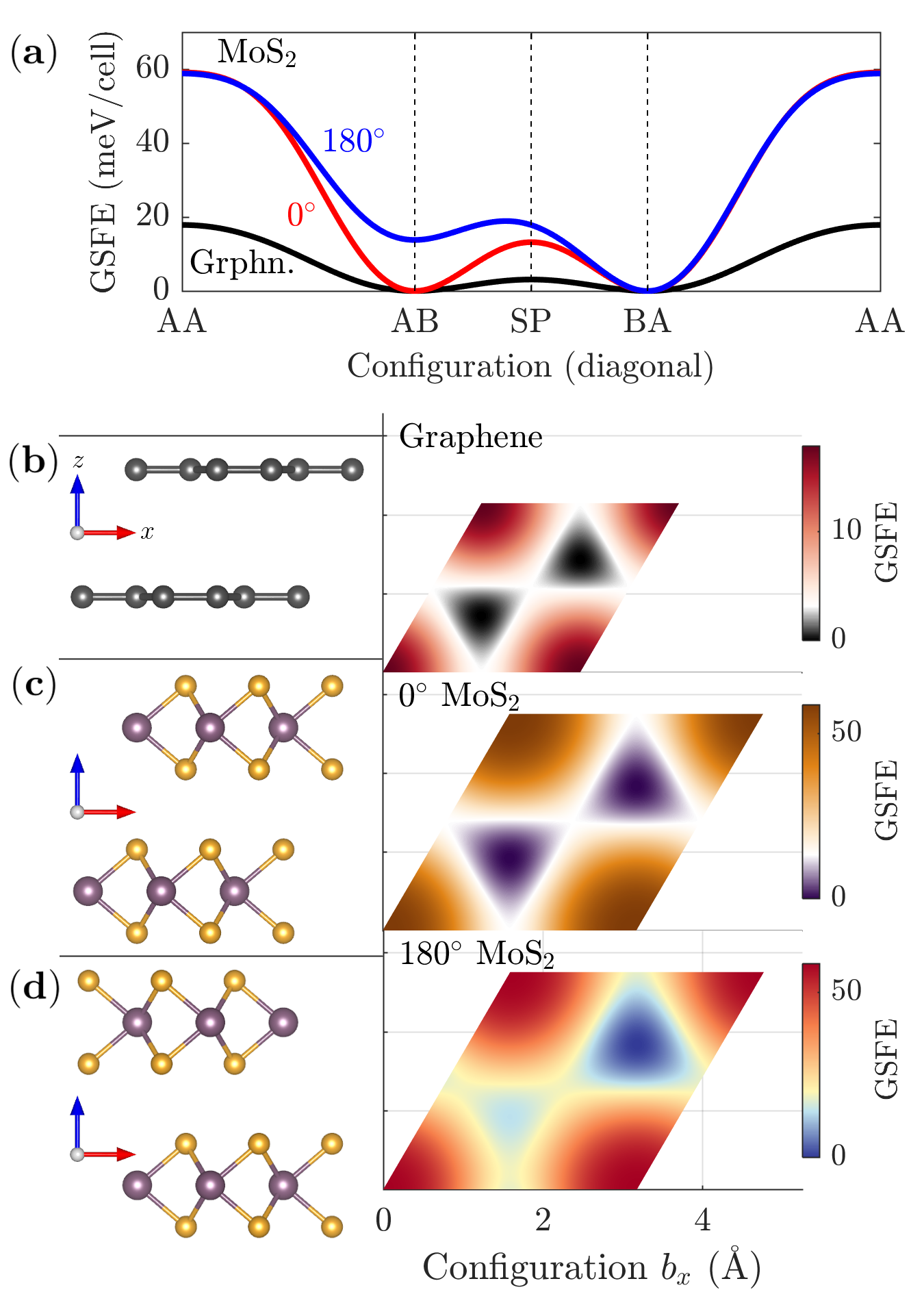}
	\caption{\textbf{(a)}: The Generalized Stacking Fault Energy (GSFE) evaluated on stackings sampled along the unit-cell diagonal for three different bilayers. \textbf{(b)},\textbf{(c)},\textbf{(d)}: Side-views of the ground-state stacking orientation for each bilayer, and corresponding GSFE dependence on configurations in the unit-cell $\Gamma$. 
	The color scale in each case is chosen to make the saddle point (SP) energy white or bright yellow.}
	\label{fig:gsfe_comparison}
\end{figure}

There is always a lowest-energy stacking between layers, and relaxation should distort the layers to maximize the area of that stacking (or stackings, in the case of degenerate ground states).
Examples of bilayers under lattice mismatches ($\lambda$) and twists ($\theta$), are displayed in  Fig. \ref{fig:lattice_alignments}.
When the lattices are nearly alligned, only small amounts of lattice straining are necessary to form a large area of uniform stacking.
As the misalignment increases the strain needed for creating uniform stacking grows larger, making domain formation less favorable.
To understand the stacking energy landscape of layered materials, we show the GSFE for graphene and the two high-symmetry stacking orientations of MoS$_2$ ($0^\circ$ and $180^\circ$ rotation between layers) in Fig. \ref{fig:gsfe_comparison}.
The two different orientations in bilayer MoS$_2$ are due to the presence of different atomic species (Mo and S) on the two sub-lattices of the honeycomb lattice.
The $0^\circ$ MoS$_2$ bilayer has two identical low-energy stackings, similar to graphene, while the $180^\circ$ MoS$_2$ bilayer has only one low-energy stacking.
Symmetry arguments can predict the critical points of the GSFE, and their relative energies can be ranked by comparing interlayer distances between atoms at each stacking.
Our use of the GSFE for interlayer interactions is expected to be more accurate than empirical atomistic potentials when they exist (for materials like graphene and hBN) \cite{Kolmogorov2005}, and allows for modeling of bilayers where no such potentials exist to our knowledge (e.g., MoS$_2$).

Although we focus on small twists slightly away from $0^\circ$ alignment, in principle the argument holds for any incommensurate angle.
Consider a commensurate angle between layers and the corresponding bilayer supercell as an effective untwisted unit cell.
The strain energy and GSFE functional for this enlarged cell can be readily obtained.
If the supercell has many atoms, the GSFE is likely very smooth and does not lead to appreciable domain formation.
However, near angles that form small supercells, such as $21.78^\circ$ for twisted honeycomb lattices, the GSFE may still have enough structure to show domain formation. We leave such investigation to future work.

Since the stacking configurations of untwisted layers gives a clear picture of the bilayer energetic landscape, framing the relaxation problem entirely in terms of configurations may prove useful.
We define $2 \times 2$ matrices $A_1$ and $A_2$ as the Bravais lattice vectors of layer 1 and layer 2, whose unit-cells are labeled as $\Gamma_1$ and $\Gamma_2$.
Any point in the Bravais lattice of layer 2 is indexed by an integer tuple, $\mathbf{n} \in \Z$, and it will have position $\mathbf{r} = A_2 \mathbf{n}$.
We can compute its stacking configuration relative to layer 1 by $\mathbf{b}_2: A_2 \Z \to \Gamma_1$ explicitly by $\mathbf{b}_2(A_2 \mathbf{n}) = A_2 \mathbf{n} = \mathbf{r}(\mathbf{n})$.
Although the function $\mathbf{b}_2$ lacks an explicit modulation in its definition, it is implicitly periodic over the torus $\Gamma_1$. 
As defined, $\mathbf{b}_2$ would vary quickly on the scale of the unitcell if $\mathbf{r}$ is formally substituted for $\mathbf{r}(\mathbf{n})$.
This is not desirable, so instead we smoothly interpolate $\mathbf{b}_2(\mathbf{r}(\mathbf{n}))$ between lattice points.
We define lattice mismatch matrices that encode the effective moir\'e pattern: $1 - A_1 A_2^{-1} = A_{\delta 2}$ and $1 - A_2 A_1^{-1} = A_{\delta 1}$, which yield interpolated mappings

\begin{equation}
\begin{split}
	\mathbf{b}_2(A_2 \mathbf{n}) &= (1 - A_1 A_2^{-1}) A_2 \mathbf{n} \\
	\implies \mathbf{b}_2(\mathbf{r}) &\equiv A_{\delta 2} \mathbf{r}, \quad \mathbf{b}_1(\mathbf{r}) \equiv A_{\delta 1} \mathbf{r}
\end{split}
\end{equation}

and a relationship between atomic displacements in configuration space, $\mathbf{u}_i(\mathbf{b}_i)$,  and those in real space:

\begin{equation}\label{eq:displacement_map}
	\mathbf{U}_i(\mathbf{r}) = \mathbf{u}_i( A_{\delta i} \mathbf{r}) \implies \nabla \mathbf{U}_i = \nabla \mathbf{u}_i A_{\delta i}.
\end{equation}

$A_{\delta i}$ can be interpreted as a map from real space to configuration space, and $A_{\delta i}^{-1}$ as the inverse map.
This transformation has been done in previous work for studying electronic structure in incommensurate materials \cite{Massatt2017, Carr2017}, and here it allows for relaxation of incommensurate bilayers.

To calculate relaxation for twisted bilayers in configuration space, the energy functionals must be defined over the configuration space $\{ \Gamma_1, \Gamma_2 \}$.
As each $\Gamma_i$ is the unit cell torus independent of twist angle, the configuration space remains periodic and compact even when the bilayer is incommensurate in real space.
Throughout the energy functionals, the real space displacement fields $\mathbf{U}_i(\mathbf{r})$ must be replaced by the configuration space displacement fields $\mathbf{u}_i(\mathbf{b}_i)$.
This substitution assumes that if two atoms on layer 2 have similar stacking relative to layer 1, then they must have similar relaxation.
For two arbitrary lattices, with the GSFE computed over $\Gamma_1$, the total energy is given as:

\begin{align} \label{eq:tot_energy}
	E_\textrm{tot}(\mathbf{u}_1, \mathbf{u}_2) &\equiv E_\textrm{inter}(\mathbf{u}_1, \mathbf{u}_2) + \sum_{i=1}^2 E_\intra^{(i)}(\nabla \mathbf{u}_i) \nonumber\\
	E_\intra^{(i)}(\nabla \mathbf{u}_i) &= \int_{\Gamma_{\tilde{i}}} \frac{1}{2} \left( \mathcal{E}_c (\nabla \mathbf{u}_i) \; C_i \; \mathcal{E}_c (\nabla \mathbf{u}_i) \right) d\mathbf{b}_{\tilde{i}}  \nonumber\\
	\mathcal{E}_c (\nabla \mathbf{u}_i) &= \frac{1}{2} \left( \nabla \mathbf{u}_i A_{\delta i} + A_{\delta i}^T \nabla \mathbf{u}_i^T \right) \\
	E_\textrm{inter}(\mathbf{u}_1,\mathbf{u}_2) &= \int_{\Gamma_1} \GSFE (\mathbf{b}_2 + \mathbf{u}_2(\mathbf{b}_2) - \mathbf{\tilde{u}}_1(\mathbf{b}_2) ) d\mathbf{b}_2 \nonumber\\
		\mathbf{\tilde{u}}_1 (\mathbf{b}_2) &\equiv \mathbf{u}_1(-A_2 A_1^{-1} \mathbf{b}_2)\nonumber
\end{align}

where $\Gamma_{\tilde{i}}$ is the unit cell of the layer opposite layer $i$. The total interlayer coupling is described by a single integral evaluation of $\GSFE$ over the configuration space of only one layer. This works well if the two layers have similar unit cells. Alternatively, the interlayer coupling can be split into two equal components of $\frac{1}{2} \GSFE$ for each $\Gamma_i$. This introduces additional complexity in the twisted case, as the relative orientations between the twisted $\Gamma_i$'s needs to be taken into account.

To illustrate how to transform the relaxation problem to configuration space, we focus on a bilayer system with small twist-angle $\theta$. Letting layer 2 be rotated counterclockwise by $\theta$ relative to layer 1 gives $A_2 = R_\theta A_1$ with $R_\theta$ the conventional rotation matrix. Then $A_{\delta 1} = 1 - R_\theta^{-1}$ and $A_{\delta 2} = 1 - R_\theta$, and we can expand the rotation matrix to first order in $\theta$ to get $\mathbf{b}_i(\mathbf{r})$:

\begin{equation} \label{eq:b_theta}
	\mathbf{b}_2(\mathbf{r}) \approx \theta
	\begin{pmatrix}
		-r_y \\
		r_x
	\end{pmatrix}
	,
	\quad
	\mathbf{b}_1(\mathbf{r}) \approx \theta
	\begin{pmatrix}
		r_y \\
		-r_x
	\end{pmatrix}.
\end{equation}

Substituting this approximation for $A_{\delta i}$ into Eq. \eqref{eq:tot_energy} also shows how the gradient of $\mathbf{u}_i$ contributes to the strain-energy $\mathcal{E}_c$ with a factor of $\theta$.
This scaling predicts that as $\theta \to 0$, eventually $\mathbf{u}_i$ can balance the intralayer and interlayer energies by forming strain solitons.
The width of the domain walls in configuration space should diminish like $\theta$, resulting in constant real space width.

\begin{figure*}[t]
	\includegraphics[width=\linewidth]{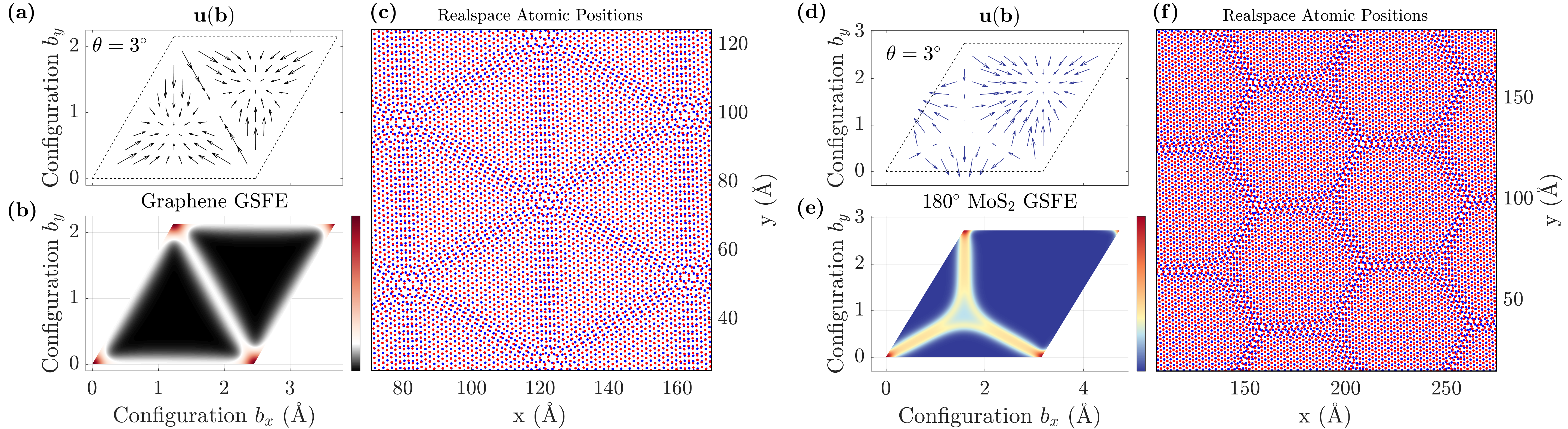}
	\caption{
	Relaxation for a graphene and $180^\circ$ MoS$_2$ bilayer with a $3^\circ$ relative twist.
	Interlayer coupling was amplified by a factor of 100 for easy visualization.
	\textbf{(a)}: The graphene displacement field $\mathbf{u}$ over $\Gamma$.
	\textbf{(b)}: $\GSFE(\mathbf{b} + 2\mathbf{u}_1(\mathbf{b}) )$ over $\Gamma$ that shows the moir\'e pattern in configuration space for graphene.
	\textbf{(c)}: The graphene atomistic positions after applying the displacement fields.
	\textbf{(d)},\textbf{(e)},\textbf{(f)} are the corresponding plots for MoS$_2$}
	\label{fig:exagerated}
\end{figure*}

\begin{figure*}[t]
	\includegraphics[width=\linewidth]{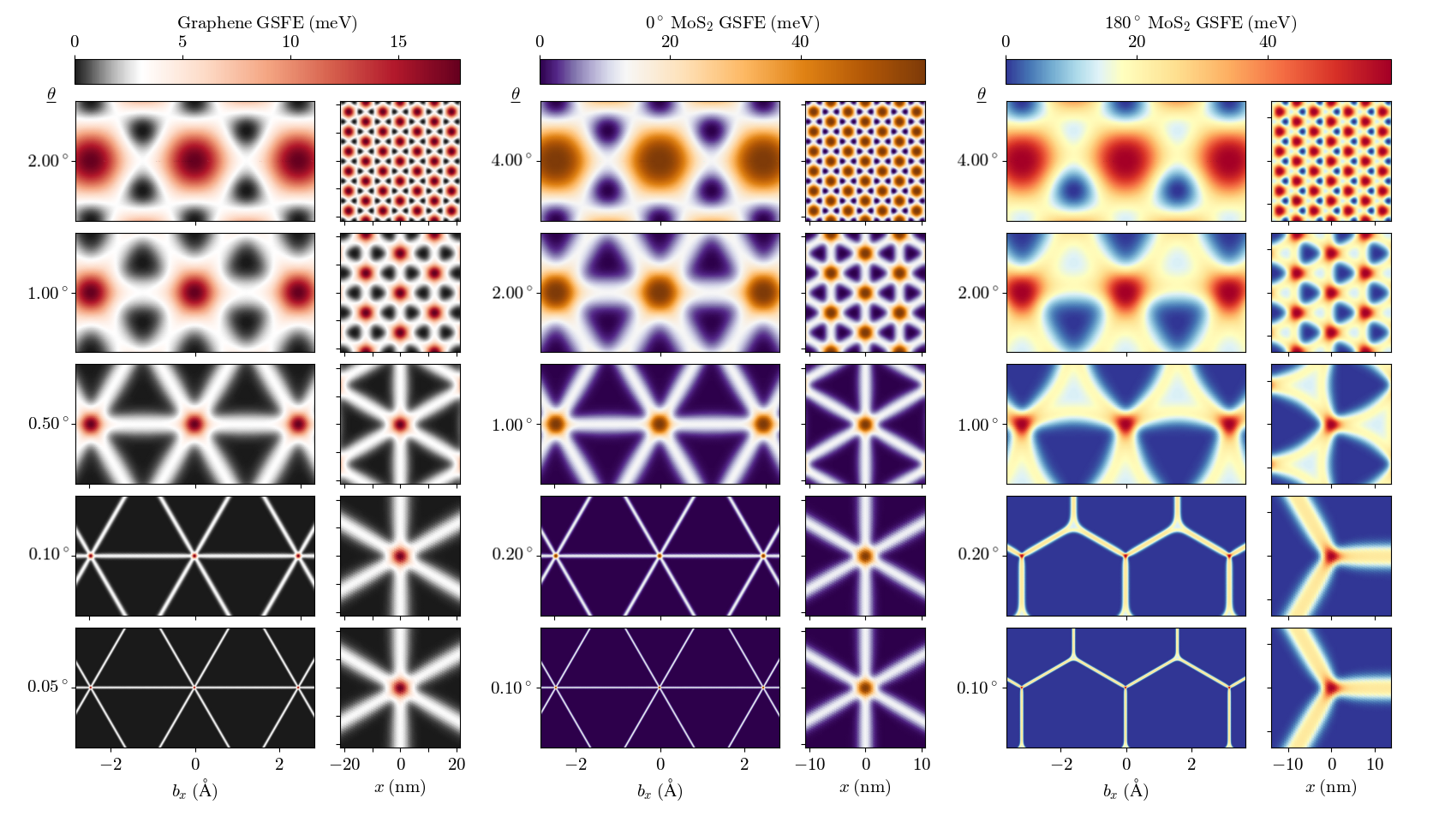}
	\caption{Configuration relaxation results for twisted bilayers with five incommensurate twist angles each. The left panel of each column shows $\GSFE(\mathbf{b} + 2 \mathbf{u}(\mathbf{b}) )$ over $\Gamma$ (the relaxation pattern in configuration space) and the right panel shows $\GSFE(\mathbf{r})$ (over real space).}
	\label{fig:relaxed_gsfe_anglesweep}
\end{figure*}

To  further simplify the model in the case of twisted bilayers, notice that $\GSFE$ depends only on the difference of atomic displacements $\Delta \mathbf{u} \equiv \mathbf{u}_2 - \mathbf{\tilde{u}}_1$.
If we consider a bilayer system where layer 1 is frozen and only layer 2 can relax, we have $\mathbf{u}_1 = 0$ and $\Delta \mathbf{u} = \mathbf{u}_2$.
As $\Delta \mathbf{u}$ minimizes the total stacking energy, when relaxing both layers we want to obtain a similar $\Delta \mathbf{u}$ while minimizing the strain energy.
The two unit cells are nearly identical, so the true solution is one that splits $\Delta \mathbf{u}$ equally between the two layers.
This can be done by setting a single $\mathbf{u}$ field $\mathbf{u} \equiv \mathbf{u}_2 = -\tilde{\mathbf{u}_1}$ over a single $\Gamma \equiv \Gamma_1$, leading to a total energy functional:

\begin{equation} \label{eq:config_E}
\begin{split}
E_\textrm{tot}(\mathbf{u}) = &\frac{1}{| \Gamma |} \int_{\Gamma} \bigg[ 2 \E_\textrm{\intra} (\nabla \mathbf{u}) + \GSFE(b + 2 \mathbf{u}) \bigg] d\mathbf{b}.
\end{split}
\end{equation}

For the three bilayers studied here, there is also a mirror-plane symmetry along the vertical plane that bisects the twist-angle, which we will label $S$. 
This symmetry gives the relation $u = S u (S b)$ as an additional constraint during any optimization procedure.
We minimize the total energy given in Eq. \eqref{eq:config_E} with a standard optimization routine implemented in the Optim Julia package\cite{Mogensen2018} after uniformly sampling configuration space with a discrete Fourier basis of plane-waves.
This yields the smooth displacement field in configuration space corresponding to the ground state of the relaxed bilayer system.
The result can then be mapped to real space for use in other applications (for example, electronic structure calculation) with Eq. \eqref{eq:displacement_map}.

To illustrate the general nature of domain formation in incommensurate graphene and TMDC bilayers, we wish to show domain-wall formation on a scale where both the unit-cell and the moir\'e supercell are easily visible.
We exaggerate the interlayer coupling by increasing $\GSFE$ by a factor of $100$ for $\theta = 3^\circ$ twisted bilayers of graphene and $180^\circ$ MoS$_2$ in Fig. \ref{fig:exagerated}.
For both systems, relaxation causes the regions of lowest energy stackings to expand in configuration space and the higher energy stackings to reduce in size, producing thin lines and nodes. 
In real space, this means the bilayers form large domains of uniform stacking surrounded by thin solitons which intersect at ``pinned'' high-energy stacking nodes.
For bilayer graphene and bilayer $0^\circ$ MoS$_2$, there are two identical ground state stackings, commonly refered to as the AB and BA stackings.
These stackings are equal in energy and compete to create a tiling of AB and BA triangular domains as observed in dark-field imaging studies of twisted bilayer graphene \cite{Alden2013, Zhang2018}.
For $180^\circ$ bilayer MoS$_2$, there is only one low-energy stacking.
It expands and causes the formation of hexagonal domains.

Furthermore, due to the anti-symmetric nature of $b(\mathbf{r})$ in Eq. \eqref{eq:b_theta}, $\nabla \times \mathbf{U}(\mathbf{r}) \approx \theta \left( \nabla \cdot \mathbf{u}(\mathbf{b}) \right)$, that is, the local change in the real space twist angle ($\nabla \times \mathbf{U}$) caused by the relaxation can be computed by taking the divergence of the configuration space displacement field.
The low-energy stackings have $\nabla \cdot \mathbf{u} < 0$, which implies an ``untwisting'' of those areas in real space.
Meanwhile, the high energy stackings have $\nabla \cdot \mathbf{u} > 0$, which implies additional twisting.
This is why the domains show almost no local twist angle in real space, while the high-energy nodes are twisted more.
We find that the local twist angle at the AA stacking in twisted bilayer graphene converges to $1.7^\circ$ as the global twist angle approaches $0^\circ$, which agrees with the results of a recent real space approach \cite{Zhang2018}.

To show that these phenomena are general, we calculate relaxed structures for various twist angles in Fig. \ref{fig:relaxed_gsfe_anglesweep}. 
We note that $\mathbf{U}_i(\mathbf{r})$ is aperiodic and captures incommensurate structure even though $\mathbf{u}(\mathbf{b})$ is periodic on $\Gamma$.
Consequently, the moir\'e domains in Fig. \ref{fig:exagerated} and \ref{fig:relaxed_gsfe_anglesweep} cannot be obtained from a supercell approach since they do not exactly repeat.
The relaxations for graphene and $0^\circ$ MoS$_2$ are almost indistinguishable, except the twist angle needed in $0^\circ$ MoS$_2$ is roughly twice what is needed in graphene for a comparable relaxed structure.
For all three structures, the strain solitons and nodes shrink in configuration space proportionally to the twist-angle.
As translating the relaxation in configuration space to real space involves a factor of $\theta^{-1}$, the shape of solitons and nodes in real space are unaffected by the twist angle at sufficiently small angles.
This is expected, as there is an optimal width for a strain soliton.
As the twist angle decreases the walls do not change in width, but only in their length as the domains become larger.

In conclusion, we have presented an approach for modeling relaxations in incommensurate systems.
The methodology, based on treating the incommensurate system consistently, has led to identification of key physical ingredients for predicting what relaxations may occur.
If the lattices are aligned close to a commensurate angle which yields a small cell, large-scale domain-wall formation is expected.
As the lattices are twisted away from such an angle, the domains will become smaller and their boundaries less sharp until almost no relaxation occurs at large misalignment.
The geometry of the domains and walls is determined by the number and nature of the critical points in the interlayer energy functional, and their domain size scales with the twist angle.
This naturally creates regular patterns of uniformly stacked bilayers divided by thin strain solitons.
If the bilayer geometry encodes important topological information for electrons, or if the strain and sharp stacking potentials act as a useful source of electron confinement, small twist angles can create regular networks of confined 1D states which are easily realized in experiment.

\begin{acknowledgements}
We acknowledge H. Yoo, R. Engelke, P. Kim, S. Fang, and K. Zhang for helpful discussions. 
Calculations were performed on the Odyssey cluster supported by the FAS Division of Science, Research Computing Group at Harvard University.
This work was supported by the ARO MURI Award No. W911NF-14-0247. 
S.B.T. is supported by the Department of Energy Computational Science Graduate Fellowship.
\end{acknowledgements}

\bibliography{relaxation}

\end{document}